\documentclass[aps,prd,preprint%twocolumn
,tightenlines,letterpaper,
nofootinbib,showpacs]{revtex4-1}
\usepackage{amssymb,latexsym}
\usepackage{amsmath,amsbsy,bbm}
\usepackage{epsfig,bm}
\usepackage{graphicx,comment}
\DeclareMathOperator{\tr}{tr}

\begin{document}
%%%%%%%%%%%%%%%%%%%%%%%%%%%%%%%%

\def\a{{\alpha}}
\def\b{{\beta}}
\def\d{{\delta}}
\def\D{{\Delta}}
\def\X{{\Xi}}
\def\e{{\varepsilon}}
\def\g{{\gamma}}
\def\G{{\Gamma}}
\def\k{{\kappa}}
\def\l{{\lambda}}
\def\L{{\Lambda}}
\def\m{{\mu}}
\def\n{{\nu}}
\def\o{{\omega}}
\def\O{{\Omega}}
\def\S{{\Sigma}}
\def\s{{\sigma}}
\def\th{{\theta}}

\def\ol#1{{\overline{#1}}}

\def\Dslash{D\hskip-0.65em /}
\def\Dtslash{\tilde{D} \hskip-0.65em /}

\def\CPT{{$\chi$PT}}
\def\QCPT{{Q$\chi$PT}}
\def\PQCPT{{PQ$\chi$PT}}
\def\tr{\text{tr}}
\def\str{\text{str}}
\def\diag{\text{diag}}
\def\order{{\mathcal O}}

\def\cF{{\mathcal F}}
\def\cS{{\mathcal S}}
\def\cC{{\mathcal C}}
\def\cB{{\mathcal B}}
\def\cT{{\mathcal T}}
\def\cQ{{\mathcal Q}}
\def\cL{{\mathcal L}}
\def\cO{{\mathcal O}}
\def\cA{{\mathcal A}}
\def\cQ{{\mathcal Q}}
\def\cR{{\mathcal R}}
\def\cH{{\mathcal H}}
\def\cW{{\mathcal W}}
\def\cM{{\mathcal M}}
\def\cD{{\mathcal D}}
\def\cN{{\mathcal N}}
\def\cP{{\mathcal P}}
\def\cK{{\mathcal K}}
\def\Qt{{\tilde{Q}}}
\def\Dt{{\tilde{D}}}
\def\St{{\tilde{\Sigma}}}
\def\cBt{{\tilde{\mathcal{B}}}}
\def\cDt{{\tilde{\mathcal{D}}}}
\def\cTt{{\tilde{\mathcal{T}}}}
\def\cMt{{\tilde{\mathcal{M}}}}
\def\At{{\tilde{A}}}
\def\cNt{{\tilde{\mathcal{N}}}}
\def\cOt{{\tilde{\mathcal{O}}}}
\def\cPt{{\tilde{\mathcal{P}}}}
\def\cI{{\mathcal{I}}}
\def\cJ{{\mathcal{J}}}

\def\eqref#1{{(\ref{#1})}}

%%%%%%%%%%%%%%%%%%%%%%%%%%%%%%%%%%%
%%%%%%%%%%%%%%%%%%%%%%%%%%%%%%%%%%%

\preprint{RBRC}

\author{Brian~C.~Tiburzi}
\email[]{btiburzi@ccny.cuny.edu}
\affiliation{
Department of Physics,
        The City College of New York,  
        New York, NY 10031, USA}
\affiliation{
Graduate School and University Center,
        The City University of New York,
        New York, NY 10016, USA}
\affiliation{
RIKEN BNL Research Center, 
        Brookhaven National Laboratory, 
        Upton, NY 11973, USA}
\date{\today}

\pacs{12.38.Gc; 12.39.Fe}

%%%%%%%%%%%%%%%%%%%%%%%%%%%%%%%%%%%%
%%%%%%%%%%%%%%%%%%%%%%%%%%%%%%%%%%%% 
\title{Chiral Corrections to Nucleon Two- and Three-Point Correlation Functions} 
%%%%%%%%%%%%%%%%%%%%%%%%%%%%%%%%%%%%
%%%%%%%%%%%%%%%%%%%%%%%%%%%%%%%%%%%%

%%%%%%%%%%%%%%%%%%%%%%%%%%%%%%%%%%%%

\begin{abstract}
We consider multi-particle contributions to nucleon two- and three-point functions 
from the perspective of chiral dynamics. 
Lattice nucleon interpolating operators, 
which have definite chiral transformation properties, 
can be mapped into chiral perturbation theory.
Using the most common of such operators, 
we determine pion-nucleon and pion-delta couplings to nucleon two- and three-point correlation functions
at leading order in the low-energy expansion. 
The couplings of pions to nucleons and deltas in two-point functions 
are consistent with simple phase-space considerations,
in accordance with the Lehmann spectral representation.
An argument based on available phase space on a torus is utilized to derive the scaling of multiple-pion couplings. 
While multi-pion states are indeed suppressed, 
this suppression scales differently with particle number compared to that in infinite volume. 
For nucleon three-point correlation functions, 
we investigate the axial-vector current at vanishing momentum transfer. 
The effect of pion-nucleon and pion-delta states on the extraction of the nucleon axial charge 
is assessed.  
We show that couplings to finite volume multi-particle states could potentially lead to overestimation of the axial charge.  
Hence pion-nucleon excited states cannot explain the trend seen in lattice QCD calculations of the nucleon axial charge. 
\end{abstract}

%%%%%%%%%%%%%%%%%%%%%%%%%%%%%%%%%%%%

\maketitle

%%%%%%%%%%%%%%%%
\section{Introduction}              %
%%%%%%%%%%%%%%%%

Dramatic progress continues to be made in strong interaction physics by solving QCD numerically utilizing Euclidean space-time lattices~%
\cite{DeGrand:2006zz}. 
Successful determination of the spectrum of low-lying hadrons from lattice QCD has been a major triumph~% 
\cite{Durr:2008zz}, 
and many recent calculations are performed at or near the physical pion mass. 
Small corrections due to strong and electromagnetic isospin breaking are becoming relevant for precision determination of some quantities, 
and the recent 
\emph{ab initio}
determination of the proton-neutron mass splitting~%
\cite{Borsanyi:2014jba}, 
for example, 
shows that lattice QCD techniques are maturing to the point required for precision low-energy QCD.   
Further benchmarks are needed for the nucleon, 
however, 
before one can reliably use the nucleon as a QCD laboratory for new physics. 
In this respect, 
determination of nucleon structure from the lattice remains a prime goal. 
For a recent overview of baryon structure from lattice QCD, 
see~%
\cite{Syritsyn:2014saa}.

The axial charge of the nucleon is one such benchmark quantity, and one for which lattice methods have expended considerable effort to determine. 
This isovector quantity does not suffer the complication of quark-disconnected contributions, 
and so arguably represents the simplest aspect of nucleon structure to compute. 
For pion masses at the physical point, 
however, 
nucleon correlation functions suffer from a well-known signal-to-noise problem~%
\cite{Lepage:1989hd}, 
which is only exacerbated in the calculation of nucleon three-point functions. 
For many years, 
lattice computations of the axial charge have been subject to underestimation, 
with a variety of systematic sources investigated for this effect, 
such as: 
finite volume~%
\cite{Jaffe:2001eb,Cohen:2001bg,Beane:2004rf,Smigielski:2007pe,Yamazaki:2008py}, 
excited-state contamination~%, 
\cite{Owen:2012ts,Capitani:2012gj,Jager:2013kha}
and even thermal effects~%
\cite{Green:2012ud}. 
Not all collaborations, 
however, 
have obtained consistent results, 
for example, 
finite volume effects could not be substantiated~%
\cite{Alexandrou:2013joa}, 
excited-state effects were not seen in the studies~%
\cite{Dinter:2011sg,Bhattacharya:2013ehc}, 
and thermal effects could not be reproduced~%
\cite{Green:2013hja}. 
The current status of such lattice QCD computations is thoroughly reviewed in~%
\cite{Constantinou:2014tga}.

We investigate the contamination of nucleon two- and three-point functions due to excited states. 
Pion-nucleon and pion-delta excited states are considered within the framework of chiral perturbation theory. 
Some time ago~%
\cite{Tiburzi:2009zp}, 
we determined chiral corrections to the time-dependence of nucleon correlation functions in infinite volume
by treating the heavy-nucleon operator of chiral perturbation theory as a model for a well-optimized lattice interpolating field. 
In fact, 
the correspondence can be made precise provided the smearing radius of the lattice operator is smaller than the pion Compton wavelength, 
see~%
\cite{Bar:2013ora,Bar:2015zwa}
for discussion of this and related points. 
This separation of scales was utilized in% 
~\cite{Bar:2015zwa}
to determine model independent pion-nucleon contributions to nucleon two-point functions. 
Chiral perturbation theory has also been used to obtain the multi-pion excited-state contributions to the axial-axial and vector-vector current correlation functions~%
\cite{Bar:2012ce}. 
In this work, 
we determine the contamination in the nucleon two-point function, and three-point function of the axial-vector current by treating chiral dynamics in finite volume. 
At leading-order in the low-energy expansion, 
the coupling of pion-nucleon and pion-delta states to these correlation functions is predicted on account of chiral symmetry. 
Such couplings to the two-point function are found to be small,%
\footnote{
This conclusion is also drawn in~%
\cite{Bar:2015zwa}.
In that work, pion-nucleon contributions to nucleon two-point functions are determined using 
relativistic baryon chiral perturbation theory. 
As we utilize a non-relativistic expansion throughout, 
$m_\pi / M_N \ll 1$, 
our results can only be compared in the non-relativistic limit, 
and in this limit all expressions indeed agree. 
} 
and those for the axial current three-point function are shown to favor overestimation of the nucleon axial charge. 
As a result, 
we conclude that pion-nucleon contamination cannot be the source of underestimation of the axial charge.

The organization of our presentation is as follows. 
First in 
Sec.~\ref{sec:nuke}, 
we detail the computation of the nucleon two-point function in chiral perturbation theory. 
One must understand how to map lattice QCD interpolating operators for the nucleon into the effective theory, 
and this is done to leading order for the standard
$(\frac{1}{2}, 0 ) \oplus ( 0, \frac{1}{2} )$
nucleon operator using heavy-nucleon chiral perturbation theory, 
Sec.~\ref{sec:NO}. 
Results for pion-nucleon and pion-delta spectral weights are obtained on a torus in 
Sec.~\ref{sec:N2}. 
From the effective mass of the two-point function calculated using chiral perturbation theory, 
we observe that the dominant pion-nucleon contamination is quite small, 
leading to at most a few percent overestimation of the nucleon mass.  
An analysis of the finite-volume couplings to multi-pion states is carried out using simple
$N$-body phase-space considerations in 
Sec.~\ref{sec:MPC}. 
On a torus, 
we find the vanishing of spectral weights at threshold, 
however, 
not as sharply as in infinite volume. 
There is one exception, 
the case of a two-body pion-nucleon state in an $s$-wave, 
for which the corresponding spectral weight does not vanish at threshold. 
Away from threshold, 
one can still argue for suppression of all multi-pion nucleon states provided 
$m_\pi L \gg 1$. 
Next we investigate the nucleon three-point function of the axial-vector current in 
Sec.~\ref{sec:ax}.
The general form of the axial-current matrix element is written out in 
Sec.~\ref{sec:ACC}, 
and expanded treating the excited-state couplings as small, 
which is the case in chiral perturbation theory. 
That calculation is carried out in Sec.~\ref{sec:CC}, 
where contributions from all possible intermediate states arising at one-loop order are enumerated. 
Chiral contamination in the axial-vector three-point function is shown in 
Sec.~\ref{sec:ACR}, 
and drives the correlation function upward, 
thus potentially leading one to overestimate the nucleon axial charge, 
$g_A$. 
A summary of findings in Sec.~\ref{sec:sum} concludes our work.

%%%%%%%%%%%%%%%%%%%%%%%%%%%%%%%%
\section{Nucleon Operator and Two-Point Function}                 %
\label{sec:nuke}
%%%%%%%%%%%%%%%%%%%%%%%%%%%%%%%%

We begin by obtaining the spectral representation of the nucleon two-point correlation function on a torus using chiral dynamics. 
To accomplish this, 
we map the standard lattice interpolating field into the corresponding nucleon operator of heavy-baryon chiral perturbation theory.  
Couplings of pion-nucleon and pion-delta states on a torus are accounted for at leading order in the 
low-energy expansion. 
We argue for the suppression of multiple-pion states using finite volume phase-space considerations.

\subsection{Nucleon Operator}
\label{sec:NO}

To study the nucleon in lattice QCD, 
one uses a quark-level interpolating field with nucleon quantum numbers, 
which we label
$\cO_N( \vec{x}, \tau)$.
To determine the nucleon mass, 
for example, 
one computes the two-point correlation function 
\begin{equation}
G(\tau)
= 
\sum_{\vec{x}} 
\langle 0 | 
\cO_N(\vec{x}, \tau) \cO_N^\dagger(\vec{0}, 0)
| 0 \rangle
.\end{equation}
Here, 
for simplicity,  
we have chosen the nucleon source location at the origin, 
and projected the correlator onto vanishing spatial momentum by summing over all lattice sites,
$\vec{x}$. 
In the limit of long Euclidean time separation,
$\tau$, 
which we have assumed is positive,   
the two-point correlation function has the expected behavior
\begin{equation}
G(\tau) = |Z_\cO|^2 e^{ - M_N \tau} + \cdots
,\end{equation}
where 
$M_N$
is the nucleon mass, 
and the
$\cdots$
represents exponentially suppressed terms. 
The coefficient of the leading exponential, 
$|Z_\cO|^2$, 
is physically the probability that the interpolating operator 
$\cO_N$
creates the nucleon amidst the tower of excited states, 
which all possess nucleon quantum numbers. 
In this work, 
we address contributions from excited states with pion-nucleon and pion-delta content.  
Although we have written the operator
$\cO_N(\vec{x}, \tau)$
with a notation that suggests locality, 
the nucleon interpolating operator need not be local.  
Generally one employs some form of gauge-invariant smearing of local operators in order to maximize overlap with the ground-state nucleon. 
As explicated in~\cite{Bar:2015zwa}, 
the coupling of pion-nucleon states to the smeared operator 
$\cO_N$
is fixed by chiral dynamics provided the smearing radius remains much smaller than the pion Compton wavelength.

Written in terms of the quark isodoublet field, 
$q = \begin{pmatrix} u \\ d \end{pmatrix}$, 
the standard local nucleon interpolating operator has the form%
\footnote{
The chiral structure of nucleon interpolating fields has been fully classified in~%
\cite{Nagata:2008zzc}.
This classification has been used in the framework of covariant baryon chiral perturbation theory to compute chiral corrections to 
moments of the nucleon distribution amplitude~%
\cite{Wein:2011ix}.
}
\begin{equation}
\cO_{N_i}
=
q_i
\left(
q^T C \tau^2 \gamma_5 q 
\right)
\label{eq:intop}
,\end{equation}
where 
$\tau^a$
are the isospin matrices.
Under a chiral transformation of the form, 
$(L, R) \in SU(2)_L \otimes SU(2)_R$, 
the quark doublet transforms  
as the direct sum of two irreducible representations, 
namely
$(\frac{1}{2}, 0) \oplus (0, \frac{1}{2} )$. 
Given the structure of 
$\cO_N$
in Eq.~\eqref{eq:intop}, 
the nucleon interpolating operator has precisely the same chiral transformation as the quark isodoublet field. 
Our goal is to find an effective field theory operator for the nucleon that shares this transformation. 
In the context of the effective theory, such an operator will be local, 
and we necessarily assume the smearing radius for the lattice interpolator is much smaller than 
$1 / m_\pi$.

In chiral perturbation theory, 
pion fields are the Goldstone modes 
emerging from spontaneous chiral symmetry breaking, 
$SU(2)_L \otimes SU(2)_R \to SU(2)_V$. 
These modes are embedded in the coset field 
$\Sigma = \exp ( 2 i \vec{\phi} \cdot \vec{\tau} / f)$, 
which transforms as 
$\Sigma \to L \Sigma R^\dagger$
under chiral transformations. 
In our conventions, 
$f = 130 \, \texttt{MeV}$, 
and the pion fields appear explicitly in the matrix
$\phi \equiv \vec{\phi} \cdot \vec{\tau}$
as
\begin{equation}
\phi 
= 
\begin{pmatrix} 
\frac{1}{\sqrt{2}} \pi^0 
& 
\pi^+ 
\\ 
\pi^- 
& 
- \frac{1}{\sqrt{2}} 
\pi^0 
\end{pmatrix}
.\end{equation} 
The universal low-energy dynamics of pions is obtained by constructing the most general chirally invariant Lagrangian density. 
This non-renormalizable theory is made tractable by utilizing a power counting that treats pion momenta 
$k$
as small compared to the chiral symmetry breaking scale, 
$\Lambda_\chi \equiv 2 \sqrt{2} \pi f$. 
Explicit chiral symmetry breaking introduced by the quark mass, 
$m_q$, 
can also be included in this power counting. 
Expanding the coset field according to this power counting, 
$\Sigma = 1 + 2 i \phi / f + \cdots$, 
we can see in the effective theory that pions are Gau\ss ian fluctuations about the chirally asymmetric vacuum.

The nucleon can be included in chiral perturbation theory as an external source of isospin which remains massive in the chiral limit. 
To maintain a low-energy expansion, 
one must take care of the fact that nucleon momenta 
$p$
can be on the order of the chiral symmetry breaking scale, 
because 
$M_N \sim \Lambda_\chi$, 
where 
$M_N$ 
is the nucleon mass. 
The solution to this problem~% 
\cite{Georgi:1990um,Jenkins:1990jv}
is to obtain a non-relativistic formulation by decomposing nucleon momentum modes as
\begin{equation}
p_\mu = M_N v_\mu + k_\mu
,\end{equation}
and integrating out anti-nucleon components of the spinor, which are 
$\sim 2 M_N$
away in energy from the nucleon components. 
The resulting heavy nucleon field,
$N_v$,
has residual momentum 
$k$, 
which satisfies the power-counting requirement
$k \ll \Lambda_\chi$. 

The form of interactions of nucleons with pions, 
for example,  
can be deduced from the fact that the nucleon transforms as a doublet under the unbroken 
$SU(2)_V$
symmetry. 
The chiral transformation properties of the nucleon field in the chiral limit are unknown, 
see~\cite{Beane:2002td} for a nice discussion and conjecture.  
The inability to resolve soft pions and nucleons in the chiral limit, 
however, 
gives one freedom to define infinitely many operators which share the same 
$SU(2)_V$
transformation, 
see~\cite{Tiburzi:2015ida} for a pedagogic overview.
The conventional choice is to utilize the field
\begin{equation}
\xi = \sqrt{\Sigma}
,\end{equation}
which has the chiral transformation
$\xi \to L \xi U^\dagger = U \xi R^\dagger$, 
where the unitary matrix 
$U$
is a function of 
$L$, 
$R$, 
and
$\xi(x)$. 
The advantage of this choice is that a nucleon operator living in any chiral multiplet can be dressed with pions using the field  
$\xi$
so that it transforms as
$N_v \to U N_v$
under 
$SU(2)_L \otimes SU(2)_R$. 
In this way, 
one can construct pion-nucleon interactions consistent with the pattern of explicit and spontaneous chiral symmetry breaking 
without knowledge of which chiral multiplet the nucleon belongs.

In the case of lattice QCD, 
nucleon interpolating operators, 
such as that in Eq.~\eqref{eq:intop},  
belong to certain chiral multiplets. 
Given the effective field theory operator
$N_v$, 
we must accordingly dress it with pions, 
so that it transforms as
$(\frac{1}{2}, 0) \oplus (0, \frac{1}{2} )$
under chiral transformations. 
That is we seek an operator 
$N$, 
which can be written in the form
$N_R + N_L$, 
where
$N_R \to R N_R$
and
$N_L \to L N_L$
under chiral transformations. 
In the low-energy expansion, 
we have 
\begin{eqnarray}
N_R
&=& 
\frac{1}{2}
\xi^\dagger N_v
+ 
\cdots
\notag \\
N_L
&=&
\frac{1}{2}
\xi N_v
+
\cdots
,\end{eqnarray}
where the 
$\cdots$
consist of operators having higher mass dimension. 
These operators appear with unfixed coefficients in the perspective of the effective theory, 
and give rise to effects beyond the order to which we are working. 
The normalization factors of 
$\frac{1}{2}$
account for the fact that the heavy nucleon field
satisfies
$\cP_{L,R} \, \Psi_N = \frac{1}{2} N_v + \cO(1/M_N)$, 
where 
$\Psi_N$
is the fully covariant nucleon spinor, 
and the chiral projectors are defined to be
$\cP_{L,R} = \frac{1}{2} ( 1 \mp \gamma_5)$.%
\footnote{
In terms of the fully covariant spinor
$\Psi_N$, 
we have the right- and left-handed fields
$N_R = \xi^\dagger \Psi_{N,R}$
and
$N_L = \xi \Psi_{N,L}$. 
These lead to a covariant nucleon interpolating operator in the effective field theory
$N_R + N_L 
= 
\frac{1}{2} \left( \xi^\dagger+ \xi \right) \Psi_N 
+ 
\frac{1}{2} \left( \xi^\dagger - \xi \right) \gamma_5 \Psi_N$, 
which is that found in%
~\cite{Bar:2015zwa}.
The $\gamma_5$-term leads to 
$\cO(1/M_N)$
effects in the non-relativistic limit, 
and these have been dropped in our work. 
}
Combining these right- and left-handed doublets into an even parity nucleon operator, 
we see
\begin{equation}
\cO_N
\longrightarrow
N
\equiv
Z_\cO
\left(
N_R
+
N_L
\right)
= 
\frac{1}{2}
Z_\cO
( \xi^\dagger + \xi )
N_v
+ 
\cdots
\label{eq:Nop}
.\end{equation}
The mapping of 
$\cO_N$
into 
$N$, 
of course, 
involves an operator-dependent coefficient
$Z_\cO$ 
that is undetermined in the effective theory, 
and depends, 
for example, 
on the quark-level smearing. 
Our results for pion-nucleon and pion-delta couplings will always be ratios to the nucleon coupling, 
for which 
$|Z_\cO|^2$
cancels out. 
That this is true can already be seen from expanding the nucleon operator 
$N$
in powers of the pion field
\begin{equation}
N
=
Z_\cO
\left(
1
-
\frac{\phi^2}{2 f^2}
\right)
N_v
+
\cdots
\label{eq:tad}
.\end{equation}
While the overall coefficient of the operator is not fixed, 
the relative coupling to two pions is fixed on account of parity and chiral symmetry.

\subsection{Nucleon Two-Point Function}
\label{sec:N2}

For the nucleon, 
we have the following Lehmann representation for the infinite volume two-point function
\begin{equation}
G(\tau)
=
|Z_\cO|^2
e^{ - M_N \tau} 
\left[
1 
+ 
\int_0^\infty dE \,
\ol \rho (E)
e^{ - E \tau}
\right]
,\end{equation}
with 
$\tau > 0$ 
assumed for the sink location. 
This is the large mass limit of the ordinary Lehmann representation, 
with the exception that we work in Euclidean space. 
The full spectral function, 
$\rho(E) = \delta(E) + \ol \rho (E)$,
includes the nucleon pole contribution to the two-point function. 
The 
residual
spectral function,
$\ol \rho (E)$, 
hence encodes the excited-state contamination. 
The long-time limit produces just the ground-state nucleon, 
with corrections being exponentially suppressed by the gap.%
\footnote{
An important distinction is that in Minkowski space, 
excited-state contributions are only power-law suppressed. 
The leading corrections arise from the accumulation of continuum states near threshold, 
where the spectral function has a cusp. 
}

Using one-loop heavy nucleon chiral perturbation theory, 
the spectral function for the nucleon operator 
$N_v$
has been obtained previously~\cite{Tiburzi:2009zp}
by evaluating the sunset diagrams shown in Fig.~1 of that work. 
This result can be written in terms of separate 
pion-nucleon and pion-delta contributions to the residual spectral function
$\ol \rho (E)$, 
namely
\begin{eqnarray}
G(\tau)
=
|Z_\cO|^2
e^{ - M_N \tau}
\left[
1 
+ 
\int_{m_\pi}^\infty
dE \, \ol \rho_{\pi N} (E) e^{ - E \tau}
+ 
\int_{m_\pi + \D}^\infty
dE \, \ol \rho_{\pi \D} (E) e^{ - E \tau}
\right]
,\end{eqnarray}
where the pion-nucleon and pion-delta spectral weights are given by
\begin{eqnarray}
\ol \rho_{\pi N} (E) 
&=&
\frac{6 g_A^2}{(4 \pi f)^2} 
\frac{[E^2 - m_\pi^2]^{3/2}}{E^2},
\notag \\
\ol \rho_{\pi \D} (E)
&=&
\frac{ 16 g_{\D N}^2}{ 3 ( 4 \pi f)^2}
\frac{[(E - \D)^2 - m_\pi^2]^{3/2}}{E^2}
\label{eq:spec}
,\end{eqnarray}
respectively.
The characteristic cusp at threshold is universal for a two-body state in a 
$p$-wave.
The gap is 
$m_\pi$ 
for the pion-nucleon continuum, 
and 
$m_\pi + \D$ 
for the pion-delta continuum, 
with 
$\D = M_\D - M_N$
as the nucleon-delta mass splitting.

%%%%%%%%%%%%%%%%%%%%%%%%%%%%%%%%%%%%%%%
%
%
%
\begin{figure}
\includegraphics[width=0.4\textwidth]{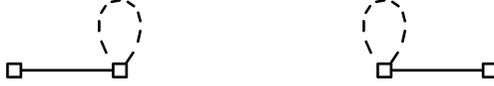}%
\caption{
Additional diagrams contributing to the nucleon two-point function beyond the sunset diagrams computed in~%
\cite{Tiburzi:2009zp}.
The squares denote the nucleon source and sink locations. 
The dashed lines represent pions, while the solid lines represent nucleons. 
}
\label{f:tad}
\end{figure}
%
%
%
%%%%%%%%%%%%%%%%%%%%%%%%%%%%%%%%%%%%%%%

To derive the spectral representation for the 
$(\frac{1}{2}, 0) \oplus (0, \frac{1}{2} )$ 
nucleon operator
$N$
in Eq.~\eqref{eq:Nop}, 
we must include the contribution from pion tadpole diagrams necessitated by the low-energy expansion of the effective field theory operator in 
Eq.~\eqref{eq:tad}. 
Evaluating the diagrams shown in Fig.~\ref{f:tad} in position space, 
however, 
we find there is no contribution to the spectral weight
because the tadpole topology produces only a renormalization of the nucleon wave-function.
When properly renormalized, 
this contribution will lead to a correction to the multi-particle couplings beyond the order we are working. 
That the tadpole topology does not contribute to the spectral weight can additionally be seen in momentum space. 
Indeed in Minkowski space, 
there is no imaginary part generated, 
and hence no multi-particle cut.

On a periodic spatial torus, 
the available multi-particle states are discrete. 
Consequently the expected form of the finite volume Lehmann representation is
\begin{equation}
G(\tau) 
= 
|Z_\cO|^2
e^{ - M_N \tau} 
\left[ 
1 
+ 
\sum_{n}
| Z_n |^2 e^{ - E_n \tau}
\right]
,\end{equation}
where 
$n$
is a collective label for the higher-lying states, 
and the energies, 
$E_n$
are measured relative to the nucleon mass. 
To determine the weights
$|Z_n|^2$, 
we must revisit the computation on a torus, 
which we take to have length 
$L$
in each of the three spatial directions. 
We keep the length of the time direction infinite throughout.

For the pion-nucleon contribution, 
we can return to an intermediate step of the computation by noting that the energy, 
$E$, 
is given by 
$E^2 = \vec{k}^2 + m_\pi^2$. 
Hence, 
we can revert the energy integration to the form of a three-momentum integral
\begin{equation}
\int_{m_\pi}^\infty
dE
= 
\int_0^\infty d |\vec{k}|  \frac{|\vec{k}|}{\sqrt{\vec{k}^2 + m_\pi^2}}
= 
2 \pi^2
\int \frac{d \vec{k}}{(2 \pi)^3} \frac{1}{| \vec{k} | \sqrt{\vec{k}^2 + m_\pi^2} }
,\end{equation}
and hence have the replacement
\begin{equation}
\int_{m_\pi}^\infty
dE
\longrightarrow
\frac{ 2 \pi^2}{L^3}
\sum_{\vec{n}}
\frac{1}{| \vec{k} | \sqrt{\vec{k}^2 + m_\pi^2} },
\label{eq:replacement}
\end{equation}
where the sum is over periodic momentum modes, 
$\vec{k} = \frac{ 2 \pi}{L} \vec{n}$, 
indexed by the mode number
$\vec{n}$. 
The energy factor in the denominator is obviously a manifestation of the Lorentz invariant measure. 
Because of the factor 
$| \vec{k} |$
appearing in the denominator, 
the two-body phase space near threshold will thus be removed from the spectral functions above when converting from a continuum to a discrete set of states. 
We will investigate the case of an 
$N$-body 
final state in finite volume below. 
For the pion-delta contributions, 
the energy is given in terms of the pion three-momentum by the relation
$( E - \D)^2 =  \vec{k}^2 + m_\pi^2$. 
These facts can be used in conjunction with Eq.~\eqref{eq:spec} to determine the finite volume couplings to pion-nucleon, 
and pion-delta states.
Using the mode numbers as labels, 
we have%
\footnote{
Our result for 
$|Z_{\vec{n}}|^2$, 
which is non-relativistic, 
agrees with the first term in the $m_\pi / M_N$ expansion of the relativistic result derived in~%
\cite{Bar:2015zwa}.
}
\begin{eqnarray}
|Z_{\vec{n}}|^2
&=&
\frac{3 g_A^2}{ 4 f^2 L^3}
\frac{\vec{k}^2}{ [ \vec{k}^2 + m_\pi^2]^{3/2}},
\notag
\\
| Z^\D_{\vec{n}}|^2
&=&
\frac{ 2 g_{\D N}^2}{ 3 f^2 L^3}
\frac{\vec{k}^2}{\sqrt{\vec{k}^2 + m_\pi^2} \left[ \sqrt{\vec{k}^2 + m_\pi^2} + \D \right]^2}
\label{eq:FVweights}
.\end{eqnarray}
As we anticipated, 
these spectral weights vanish with 
$\vec{k}^2$
unlike their continuum counterparts, 
which vanish as
$|\vec{k}|^3$. 
Moreover the dependence is now analytic near threshold.

%%%%%%%%%%%%%%%%%%%%%%%%%%%%%%%%%%%%%%%
%
%
%
\begin{figure}
\includegraphics[width=0.45\textwidth]{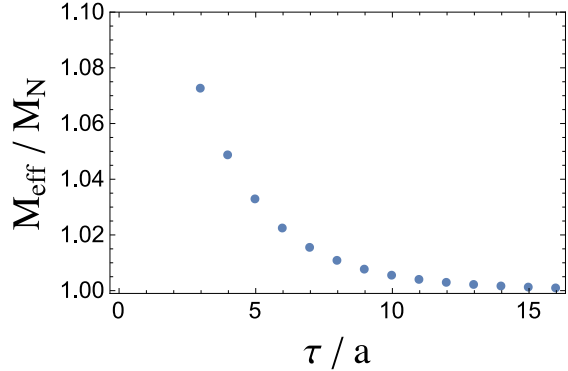}%
\caption{
Effective mass of the nucleon two-point function computed with chiral perturbation theory, 
Eq.~\eqref{eq:ChPT2}. 
Pion-nucleon excited-state contamination is shown at the few-percent level for typical values of lattice parameters. 
For times such that
$\tau > 5$, 
the correlator is saturated by pion-nucleon states with momentum mode numbers satisfying
$\sqrt{\vec{n}^2} < 4$. 
}
\label{f:effm}
\end{figure}
%
%
%
%%%%%%%%%%%%%%%%%%%%%%%%%%%%%%%%%%%%%%%

We can now write the resulting nucleon two-point function in finite volume as
\begin{equation}
G(\tau)
=
| Z_\cO |^2 
e^{ - M_N \tau}
\left[
1
+ 
\sum_{\vec{n}}
\Big(
| Z_{\vec{n}}|^2
e^{ - E_{\vec{n}} \tau}
+
| Z^\D_{\vec{n}}|^2
e^{ - (E_{\vec{n}} + \D) \tau}
\Big)
\right].
\label{eq:ChPT2}
\end{equation}
In the above expression, 
$E_{\vec{n}}$
is the pion energy. 
As a result, 
the energy appearing in the second exponential, 
$E_{\vec{n}} + M_N+ \Delta = E_{\vec{n}} + M_\Delta$, 
is the total energy of the pion-delta intermediate state in the non-relativistic limit. 
Due to the additional exponential suppression, 
$e^{- \Delta \tau}$, 
contributions from delta intermediate states are quite small. 
The pion-delta spectral weights,
however, 
are needed below to calculate the three-point function, 
for which delta contributions are non-negligible. 
To assess the size of pion-nucleon contamination in the two-point function, 
we form the effective mass of the chiral perturbation theory correlator
\begin{equation}
M_\text{eff} (\tau)
= 
- \log \frac{G(\tau + a)}{G(\tau)}
,\end{equation}
where 
$a$
is the lattice spacing. 
The effective mass is shown in Fig.~\ref{f:effm}, 
where we have employed the lattice parameters: 
$a = 0.1 \, \texttt{fm}$, 
and
$L = 48 \, a$.
We have chosen to plot results at the physical value of the pion mass, 
using the couplings 
$g_A = 1.25$
and
$g_{\D N} = 1.5$,  
along with the nucleon-delta mass splitting
$\D = 0.29 \, \texttt{GeV}$. 
The pion-delta contributions, 
however,
cannot be discerned in the plot. 
Pion-nucleon contamination is seen to be at the level of a few percent. 
The smallness of such contributions owes to three facts. 
\begin{enumerate}
\item
Pion-nucleon interactions are relatively weak, 
which is exhibited by the factor 
$(f L)^{-2}$
in the spectral weights. 

\item
Lowering the pion mass at fixed lattice size reduces the energy gap, however, it also brings one closer to threshold, 
for which the spectral weights vanish.  

\item
Raising the lattice volume at fixed pion mass also reduces the energy gap, but similarly brings one closer to threshold. 
 
\end{enumerate}
Despite the fact that spectral weights vanish less rapidly near threshold compared to the case of a continuum of states, 
the suppression at threshold in finite volume is enough to ensure pion-nucleon contamination is likely under control 
for compact lattice interpolating operators.

%%%%%%%%%%%%%%%%%%%%%
\subsection{Multi-Particle Contributions}%
\label{sec:MPC}
%%%%%%%%%%%%%%%%%%%%%

For a continuum of states,  
one can derive the behavior of the spectral function near an $N$-particle threshold~%
\cite{Brown:1992db}. 
Let the energy of the 
$N$-particle state be 
$E_{\text{Th}} = \sum_i^N M_i$, 
where the 
$M_i$
are masses of the individual particles. 
The spectral function then has the behavior%
\footnote{
There can be additional suppression by integer powers of the available energy
due to angular momentum considerations. 
For a state of orbital angular momentum
$\ell$, 
we have an additional factor of 
$( E - E_{\text{Th}})^\ell$.
}
\begin{equation}
\rho_N (E)
\propto
\sqrt{ E - E_{\text{Th}}}^{ \, 3 N - 5}
,\end{equation}
for 
$E \gtrsim E_{\text{Th}}$. 
From our discussion above, 
we know this characteristic factor arises from the integrals over momentum space, 
and so must be reconsidered on a torus.

In the derivation of the continuum result, 
the starting point is the multi-particle phase space integration
\begin{equation}
\rho_N (E)
= 
\prod_{i=1}^N 
\int \frac{d \vec{k}_i}{2 E_{\vec{k}_i} (2 \pi)^3}
(2 \pi)^4
\delta^{(4)} 
\left( 
\sum_{j=1}^N  (k_{j})_\mu - P_\mu 
\right)
| \mathcal{Z}_N |^2
\label{eq:Nphase}
,\end{equation}
where the energy of the $i$-th particle is 
$E_{\vec{k}_i} = \sqrt{ \vec{k} {}_i^2 + M_i^2}$, 
and 
$\mathcal{Z}_N$
is an overlap factor can depend on the particle momenta. 
While the general result is Lorentz invariant, 
for simplicity we work in the rest frame, 
where 
$\vec{P} = \vec{0}$. 
For energy just above threshold, 
$E \gtrsim E_{\text{Th}}$, 
all of the individual particles are non-relativistic. 
This fact can be used to treat the 
overlap factors as constants, 
which is fine provided there is no orbital angular momentum.  
The $N$-th particle's momentum, 
say, 
is entirely fixed by three-momentum conservation. 
The energy conserving delta-function puts a constraint on the magnitude of the
$(N-1)$-st 
particle's momentum in the form
\begin{equation}
\int dk_{N-1} \,
k_{N-1}^2 \,
\delta
\left(
\sum_{i=1}^N E_{\vec{k}_i}
-
E
\right)
\sim
\int dk_{N-1} \,
k_{N-1} \,
\delta
\Bigg(
k_{N-1}
-
\sqrt{2 M_{N-1} \left(E - E_{\text{Th}}\right)}
\Bigg)
.\end{equation}
To bound the 
$N-2$ 
momentum integrals, 
we note that the maximum of any remaining particles' energy is set by 
$E - E_{\text{Th}}$. 
Hence we have
\begin{equation}
\rho_N (E)
\sim
\sqrt{E - E_{\text{Th}}} \,
\prod_{i=1}^{N-2} 
\int_0^{E - E_{\text{Th}}}
d (k_i)^2 
\sqrt{ k_{i}^2}  
\propto
\sqrt{  E - E_{\text{Th}} }^{\, 3 N - 5} 
,\end{equation}
which is the known scaling.

For a torus, 
the starting point instead of 
Eq.~\eqref{eq:Nphase}
is
\begin{equation}
|Z_N (E)|^2
=
\prod_{i=1}^N 
\frac{1}{L^3}
\sum_{\vec{n}_i = - \infty}^\infty
\frac{1}{2 E_{\vec{k}_i}}
L^3 \delta^{(3)}_{\vec{0}, \sum_{j=1}^N \vec{n}_j}
\delta \, {}_{E, \sum_{j=1}^N E_{\vec{k}_j}}
| \mathcal{Z}_N |^2
\label{eq:NphaseFV}
,\end{equation}
where 
$E$ 
is a non-integer index characterizing the energy of the 
$N$-particle 
state.
Notice in writing the energy-conserving Kronecker delta in 
Eq.~\eqref{eq:NphaseFV}, 
we have the total energy as the sum of the single particle energies.
This is an approximation valid for weak interactions among the particles, 
because interactions modify the finite volume spectrum of multi-particle states.  
The weight factor 
$|Z_N(E)|^2$
is dimensionless because it is to be summed over the possible values for
$E$, 
while 
$\rho_N(E)$
has inverse mass dimension because it is to be integrated as a function of 
$E$. 
The case 
$N = 2$ 
appears to be somewhat special, 
because it can be evaluated more generally than at threshold. 
For 
$N=2$, 
we have
\begin{equation}
|Z_2(E)|^2
\approx
\frac{1}{2 M L^3}
\sum_{\vec{n}=-\infty}^\infty
\frac{1}{
2 E_{\vec{n}}}
\delta_{E, M+ E_{\vec{n}}}
| \mathcal{Z}_2 |^2
,\end{equation}
where we treat one particle of mass
$m$
as light,
and the other particle of mass
$M$
as heavy. 
We have not appealed to the threshold condition here. 
Because the overlap factor can only depend on 
$\vec{n}^2$, 
the sum collapses to a fixed value of 
$\vec{n}^2$
determined by the size of 
$E$, 
and we find
\begin{equation}
|Z_2(E)|^2
\propto
\frac{| \mathcal{Z}_2 |^2}{4 M L^3 \sqrt{\vec{k}^2 + m^2} } 
\label{eq:FV2part}
,\end{equation}
where the total energy is 
$E \approx M + \sqrt{\vec{k}^2 + m^2}$. 
Unless there are momentum factors due to orbital angular momentum,  
no suppression is found near threshold.
This occurs, 
for example, 
in the case of pion-nucleon coupled to an 
$s$-wave.   
For relative 
$p$-waves, 
our chiral perturbation theory result nicely lines-up with this general two-body analysis.
In the low-energy expansion, 
$| \mathcal{Z}_2 |^2 \propto \frac{ \vec{k}^2}{\vec{k}^2 + m_\pi^2}$, 
due to the orbital angular momentum.

Having handled the special two-body case, 
we now proceed to analyze the finite volume phase space with generic particle number
$N \geq 2$.
Considering specifically near threshold, 
all particles are non-relativistic, 
and we choose the $N$-th particle's three momentum in Eq.~\eqref{eq:NphaseFV} to vanish.  
Each sum over triplets of integers, 
$\vec{n}$,
can be changed to a sum over a radial integer
$n$, 
with a multiplicity weight 
$f(n)$
that counts the number of triplets 
$\vec{n}$ 
that can sum to a given
$n = \sqrt{ \vec{n}^2 }$.
For large 
$n$, 
we approach a continuum of states, 
and 
$f(n) \propto n$. 
For small $n$,
we take
$f(n) \sim 1$ 
because corrections to this will give additional suppression at threshold.
The remaining constraint due to the energy conserving Kronecker delta
fixes the value of the 
$(N-1)$-st 
particle's energy. 
There is no change of variables factor here, 
however, 
we still have the multiplicity weight
$f(n_{N-1})$, 
with 
$n_{N-1}$
fixed by energy conservation. 
Thus we are led to
\begin{equation}
|Z_N(E)|^2
\sim
f\Bigg( 
\frac{L}{2\pi}
\sqrt{
2 M_{N-1} 
(E - E_{\text{Th}})}
\Bigg)
\prod_{i=1}^{N-2}
\sum_{n_i= 0}^{\sqrt{E - E_{\text{Th}}}}
f(n_i) 
\propto
\sqrt{E - E_{\text{Th}}}^{\, N - 2}
,
\label{eq:FVthresh}
\end{equation}
where we treat 
$M_{N-1} L$
of generic size, 
but consider
$(E - E_{\text{Th}}) L \ll 1$. 
From the intermediate expression, 
we can take the limit of large mode numbers and recover the correct scaling for continuum states. 
To achieve this, 
notice the behavior
$\Delta n / n \to dn$, 
$f(n) \to n$, 
and we must accordingly take 
$(E - E_{\text{Th}}) L \gg1$
in the infinite volume limit. 
For multi-particle states greater than two-body, 
we find the finite volume result, 
Eq.~\eqref{eq:FVthresh}, 
maintains the vanishing of spectral weights near threshold, 
but not as sharply as for continuum states.

For the case of a nucleon with pion interactions constrained by chiral dynamics, 
we expect the spectral weights to behave as
\begin{equation}
|Z_N(E)|^2
\sim
\left(
\frac{g_A^2}{E_{\vec{k}} f^2 L^3}
\right)^{N-1}
\left(
\frac{ |\vec{k} | }{E_{\vec{k}}}
\right)^{2 ( N + \ell - 2)}
,\end{equation}
with 
$E - E_{\text{Th}} \approx \frac{\vec{k}^2}{2 m_\pi} $.  
Away from threshold, 
the dimensionless ratio
$|\vec{k} | / E_{\vec{k}}$
is replaced by some complicated function of this ratio,
$F ( |\vec{k} | / E_{\vec{k}} )$.
The pre-factors can still be used to argue suppression of multi-particle states in the general case, 
provided that
$f L \gtrsim 1$, 
and 
%$E_{\vec{k}} L > 
$m_\pi L \gg 1$. 
Contamination from multiple pions in the nucleon two-point function should thus be largely irrelevant provided the smearing radius of the interpolating operator is small compared to the pion Compton wavelength.

%%%%%%%%%%%%%%%%%%%%%%%
\section{Axial Current Three-Point Function}%
\label{sec:ax}
%%%%%%%%%%%%%%%%%%%%%%%

Having derived the pion-nucleon and pion-delta contributions to the nucleon two-point function, 
we now turn our attention to the determination of the nucleon three-point function of the axial-vector current. 
The case of three-point functions is complicated by the lack of a spectral representation. 
As a result, 
couplings to various multi-particle states have signs determined by the underlying dynamics. 
Cancellations can exist between these contributions, 
and we investigate whether chiral dynamics in finite volume drives the axial three-point function above or below the single-nucleon contribution. 
In infinite volume, 
multi-particle couplings drive the axial three-point function above the nucleon contribution~%
\cite{Tiburzi:2009zp}, 
which, 
however, 
would lead to an overestimation of the axial charge, 
$g_A$.
The same is true for discrete multi-particle states in finite volume.

%%%%%%%%%%%%%%%%%%%%
\subsection{Axial Current Correlator}%
\label{sec:ACC}
%%%%%%%%%%%%%%%%%%%%

Our interest lies in the three-point function formed from the iso-vector axial-vector current
inserted between two single-nucleon states. 
To determine the axial charge of the nucleon, 
we project the initial and final states onto vanishing three-momentum. 
Writing the current as
$J^+_{5\mu} (\vec{y}, t)$, 
we have the un-amputated three-point function
\begin{equation}
G_{5 \mu}(\tau,t)
=
\sum_{\vec{x},\vec{y}}
\langle 0 | 
\cO_N(\vec{x}, \tau) J^+_{5\mu} (\vec{y}, t) \cO_N^\dagger (\vec{0}, 0) 
| 0 \rangle
.\end{equation}
This correlation function depends on the source-sink separation,
$\tau$,
as well as the current insertion time,
$t$,
which we have assumed satisfy the hierarchy 
$\tau > t > 0$. 
Ideally one works in the limit in which 
$\tau \gg t \gg 0$, 
so that single-nucleon states can be cleanly isolated from both the source and sink.

Inserting complete sets of hadronic states into the three-point function, 
we have the general expression
\begin{eqnarray}
G_{5 \mu} (\tau,t)
&=&
| Z_\cO |^2
e^{ - M_N \tau}
\sum_{m,n} 
e^{ - E_n ( \tau - t) - E_m t } 
Z_n Z_m^*
\langle n | J^+_{5 \mu} | m \rangle
.\end{eqnarray}
Notice for ease below, 
we have factored out an overall exponential involving the nucleon mass and an overall operator overlap factor 
$| Z_\cO |^2$.
Thus all of the energies are differences relative to 
$M_N$, 
and all of the amplitudes 
$Z_n$
are ratios relative to 
$|Z_\cO|$. 
The ground-state contribution from the nucleon is isolated at long Euclidean time separations
\begin{eqnarray}
G_{5 \mu} (\tau,t)
&\overset{\tau \gg t \gg 0}{=}&
| Z_\cO |^2 
e^{ -M_N \tau} 
\langle N (\vec{0} \,) | J^+_{5 \mu} | N ( \vec{0} \, ) \rangle
+ 
\cdots,
\end{eqnarray}
where exponentially suppressed contributions from excited states have been dropped. 
In a non-relativistic notation, 
the axial-current matrix element in the nucleon has the form
\begin{equation}
\langle N (\vec{0} \,) | J^+_{5 \mu} | N ( \vec{0} \, ) \rangle
=
2 g_A \,
u^\dagger S_\mu u
,\end{equation}
where 
$u$ 
is a rest-frame spinor, 
and 
$S_\mu$
is the covariant spin operator. 
The constant
$g_A$
is the axial charge of the nucleon in the standard normalization. 

Considering now the general transition matrix elements of the axial current, 
we can write them in the form
\begin{equation}
\langle n | J^+_{5 \mu} | m \rangle =  g_{nm} \, e^{ i \varphi_{nm}} \, 2 u^\dagger S_\mu u,
\end{equation} 
with 
$g_{nm}$
as 
real-valued couplings, 
and 
$\varphi_{nm}$
as phases.
The spinors appearing above characterize the multi-particle states.
These are, 
in turn, 
constrained by the nucleon interpolating operator. 
The only overlap factors, 
$Z_n$, 
that are non-vanishing correspond to states 
$| n \rangle$
which have vanishing total three momentum, 
and total angular momentum of 
$\frac{1}{2}$. 
Such a state can be described by a non-relativistic Pauli spinor 
$u$
that merely characterizes whether the total angular momentum is aligned or anti-aligned with respect to some axis. 
Hence we define the transition matrix elements using the common spinor product 
$2 u^\dagger S_\mu u$, 
keeping in mind that any differing normalization factors for multi-particle states are thereby absorbed in the 
$g_{nm}$
couplings.

On account of Hermiticiy of the axial-vector current
and time-reversal invariance, 
we have symmetric couplings, 
$g_{nm} = g_{mn}$, 
and 
antisymmetric phases
$\varphi_{nm} = - \varphi_{mn}$. 
The latter condition implies that the imaginary part of the correlator vanishes. 
Taking this into account, 
we have the general form of the axial-three point function
\begin{eqnarray}
G_{5\mu}(\tau,t)
&=&
2 u^\dagger S_\mu u \,
|Z_\cO|^2
e^{ - M_N \tau}
\sum_{n,m} 
g_{nm} \cos \varphi_{nm}
|Z_n| |Z_m|  
e^{ - E_n  (\tau - t)} 
e^{ - E_m t} 
\label{eq:axcor}
\end{eqnarray}
To analyze excited-state contributions to this correlator, 
we amputate the external legs by forming the standard ratio of three-point to two-point functions
\begin{equation}
R_{5 \mu}
(\tau, t) 
= 
\frac{G_{5 \mu}(\tau,t)}{G(\tau)} 
.\end{equation}
Isolating the single-nucleon contribution, 
this ratio has the general form
\begin{equation}
R_{5 \mu}
(\tau, t) 
= 
2 u^\dagger S_\mu u
\left[
g_{A} + G_A(\tau,t)
\right]
.\end{equation}
Notice the ground-state to ground-state coupling
$g_{00}$
is merely the nucleon axial charge
$g_{00} = g_A$. 
While the single nucleon probability
$|Z_\cO|^2$
is not unity, 
it cancels out of the ground-state contribution in the ratio of correlators. 
The time-dependent function, 
$G_A(\tau,t)$,
thus characterizes the excited-state contamination; 
and, 
on account of the general form for the correlator given in Eq.~\eqref{eq:axcor},
it appears as
\begin{eqnarray} 
G_A(\tau,t)
&=&
\sum_{n>0}  
\left(
g_{nn} 
- g_{00} \right) 
|Z_n|^2 e^{ - E_n \tau} 
+
\sum_{n \neq m} 
g_{nm} \cos \varphi_{nm}
|Z_n| |Z_m| 
e^{ - E_n ( \tau - t ) }
e^{ - E_m t}
\label{eq:3GEN}
.\end{eqnarray}
To arrive at this form, 
we made the approximation of neglecting any terms with four or more powers of overlap factors 
$|Z_n|$.
Such terms will be higher order in the chiral expansion, 
and arise from expanding out excited-state contributions to the two-point function in the denominator. 
The leading excited-state contribution from the denominator, 
however, 
gives rise to the 
$(- g_{00})$-piece present in the diagonal sum. 
This subtracted term is analogous to wave-function renormalization.

In order to determine the multi-particle couplings, 
we find it efficacious to rearrange Eq.~\eqref{eq:3GEN} slightly. 
With redefined couplings, 
we have the compact expression
\begin{eqnarray} 
G_A(\tau,t)
&=&
\sum_{n,m - \{ 0, 0 \}} 
\mathfrak{g}_{nm} 
|Z_n| |Z_m| 
e^{ - E_n ( \tau - t ) }
e^{ - E_m t}
\label{eq:GEN}
,\end{eqnarray}
for which the relation between couplings is
\begin{equation}
\mathfrak{g}_{nm} 
= 
\delta_{nm}
(g_{nn} - g_A)
+
( 1 - \delta_{nm} )
g_{nm} \cos \varphi_{nm}
.\end{equation}

%%%%%%%%%%%%%%%%%%%%
\subsection{Chiral Computation}         %
\label{sec:CC}
%%%%%%%%%%%%%%%%%%%%

One can determine the excited-state contamination in finite volume using chiral dynamics. 
Previously the leading-order expression for the contamination function 
$G_A(\tau, t)$
was obtained for the nucleon operator 
$N_v$
in infinite volume~%
\cite{Tiburzi:2009zp}. 
See Fig.~3
of that work for the corresponding Feynman diagrams. 
After some rewriting, 
the terms appearing in that result can be grouped according to the time dependence of the general expression given above in Eq.~\eqref{eq:GEN}. 
To match onto this expression, 
we must additionally convert from the continuum of states used previously to the discrete states permitted on a spatial torus. 
Finally we must compute pion tadpole diagrams that result from the effective theory representation of the 
$(\frac{1}{2}, 0) \oplus (0, \frac{1}{2} )$ 
lattice interpolating field in Eq.~\eqref{eq:tad}. 
These diagrams account for the fact that the effective theory nucleon operator 
$N$
is perturbatively close to 
$N_v$.
The required diagrams differ from those appearing in Fig.~\ref{f:tad} merely by the insertion of an axial-current interaction along the nucleon line. 
It is a straightforward exercise to see that such axial-current operator tadpoles cancel against the tadpoles in Fig.~\ref{f:tad} when computing the ratio of three-point to two-point functions. 
This is not surprising, 
as the latter tadpoles contribute only to the wave-function renormalization. 
It thus remains to dissect the one-loop computation to obtain the multi-particle couplings.

%%%%%%%%%%%%%%%%%%%%%%%%%%%%%%%%%
\subsubsection{Nucleon to Excited State Matrix Elements}
%%%%%%%%%%%%%%%%%%%%%%%%%%%%%%%%%

The axial-current correlation function receives contributions from intermediate nucleon-to-excited state transitions. 
Because the overall nucleon mass dependence has been factored out of the three-point to two-point function ratio, 
these nucleon transition matrix elements are easily identified, 
because they lack dependence on either the source to current insertion separation, 
$t$, 
or current insertion to sink separation, 
$\tau - t$. 
First, 
we handle the matrix elements where the excited state is a pion-nucleon state.  
In infinite volume, 
the contribution from nucleon to pion-nucleon matrix elements can be written simply in terms of the residual spectral functions.
We have
\begin{eqnarray}
G_A^{N,\pi N} (\tau, t) 
&=&
\int_{m_\pi}^\infty
dE
\left(
\frac{8}{9} g_A \,
 \ol \rho_{\pi N} (E)
- 
\frac{16 g_{\D N}}{27 \sqrt{2}}
\sqrt{\ol \rho_{\pi N} (E)
\ol \rho_{\pi \D} (E+ \D)
}
\right)
\left[
e^{- E ( \tau - t)} + e^{ - E t}
\right]
.\notag \\
\end{eqnarray}
This expression can be reverted to an integral over the pion's three momentum and this integral replaced by a momentum mode sum, 
as in 
Eq.~\eqref{eq:replacement}. 
Carrying out these steps results in
\begin{eqnarray}
G_A^{N,\pi N} (\tau, t) 
&=&
\sum_{\vec{n}} 
\left(
\frac{8}{9} g_A
| Z_{\vec{n}}|^2
- 
\frac{16 g_{\D N}}{27 \sqrt{2}}
|Z_{\vec{n}}| \, |Z_{\vec{n}}^\D|
\right)
\left[ 
e^{- E_{\vec{n}} ( \tau - t)} 
+
 e^{ - E_{\vec{n}} t} 
\right]
,\end{eqnarray}
in finite volume. 
The weight factors
$|Z_{\vec{n}}|$, 
and
$|Z_{\vec{n}}^\D|$
have been determined from the nucleon two-point function calculation, 
and appear in Eq.~\eqref{eq:FVweights}. 
Comparing with Eq.~\eqref{eq:GEN}, 
we see the first exponential above corresponds to contributions from 
$N$-to-$\pi N$ 
transition matrix elements, 
whereas the second are symmetrical 
$\pi N$-to-$N$
matrix elements. 
Given our definition of multi-particle couplings, 
we find that
\begin{eqnarray}
\mathfrak{g}_{N, \pi(\vec{n}) N} &=& \frac{8}{9} g_A |Z_{\vec{n}}|
- 
\frac{16 g_{\D N}}{27 \sqrt{2}} 
|Z_{\vec{n}}^\D|,
\end{eqnarray}
where we necessarily label the pion-nucleon state by the pion momentum mode numbers.

In infinite volume, 
intermediate state nucleon to pion-delta matrix elements can also be readily identified by their time dependence. 
We have the contributions
\begin{eqnarray}
G_A^{N,\pi \D} (\tau, t) 
&=&
\left( g_A + \frac{25}{81} g_{\D\D} \right)
\int_{m_\pi + \D}^\infty dE \, 
\ol \rho_{\pi \D} (E) 
\left[
e^{- E ( \tau - t)} + e^{ - E t} 
\right]
\notag \\
&&
- 
\frac{16 g_{\D N}}{27 \sqrt{2}}
\int_{m_\pi}^\infty 
dE
\sqrt{\ol \rho_{\pi N} (E)
\ol \rho_{\pi \D} (E+ \D)
}
\left[ 
e^{ - (E + \Delta) ( \tau - t)} + e^{ - (E + \Delta) t} 
\right]
,\end{eqnarray}
which, 
on account of 
Eq.~\eqref{eq:replacement},
can be cast in the form
\begin{eqnarray}
G_A^{N,\pi \D} (\tau, t) 
&=&
\sum_{\vec{n}} 
\left[
\left( g_A + \frac{25}{81} g_{\D\D} \right)
| Z^\D_{\vec{n}}|^2 
- 
\frac{16 g_{\D N}}{27 \sqrt{2}}
|Z_{\vec{n}}| \, |Z_{\vec{n}}^\D|
\right]
\left[ 
e^{- (E_{\vec{n}} + \D) ( \tau - t)} 
+
 e^{ - (E_{\vec{n}} + \D)t} 
 \right]
,\notag \\
\end{eqnarray}
in finite volume. 
Comparing with Eq.~\eqref{eq:GEN}, 
we see the first exponential above corresponds to contributions from 
$N$-to-$\pi \D$ 
transition matrix elements, 
whereas the second contributions are symmetrical 
$\pi \D$-to-$N$
matrix elements. 
Given our definition of multi-particle couplings, 
we find that
\begin{eqnarray}
\mathfrak{g}_{N, \pi(\vec{n}) \D} 
&=& 
\left( g_A +  \frac{25}{81} g_{\D \D} \right) |Z^\D_{\vec{n}}|
- 
\frac{16 g_{\D N}}{27 \sqrt{2}}
|Z_{\vec{n}}| 
,\end{eqnarray}
where pion-delta states have been labeled by the pion momentum mode numbers. 
This exhausts all nucleon to excited state matrix elements appearing in the axial-current correlation function. 
Before continuing, 
we remark that all multi-particle contributions to the three-point function above survive when either 
$t \gg 0$
or 
$\tau \gg t$, 
but not both.

%%%%%%%%%%%%%%%%%%%%%%%%%%%%%%%%%
\subsubsection{Excited State to Excited State Matrix Elements}
%%%%%%%%%%%%%%%%%%%%%%%%%%%%%%%%%

The remaining contributions to the three-point correlation function arise from 
intermediate excited state to excited state matrix elements of the axial current. 
There are contributions from both 
$\pi N$-to-$\pi N$ 
matrix elements, 
as well as 
$\pi \Delta$-to-$\pi \D$
matrix elements. 
In infinite volume, 
these such contributions appear in the expression 
\begin{eqnarray}
G^{\pi B, \pi B}_A (\tau, t)
&=&
- \frac{8}{9} g_A
\int_{m_\pi}^\infty
dE \, 
\ol \rho_{\pi N} (E)
\, e^{ - E \tau}
-
\left( 
g_A + \frac{25}{81} g_{\D \D}
\right)
\int_{m_\pi + \D}^\infty 
dE \, 
\ol \rho_{\pi \D}(E) \, 
e^{ - E \tau}
.\end{eqnarray}
Using 
Eq.~\eqref{eq:replacement}, 
the corresponding expression on a spatial torus appears in the form
\begin{eqnarray}
G^{\pi B, \pi B}_A (\tau, t)
&=&
- \sum_{\vec{n}}
\left[
\frac{8}{9} g_A
|Z_{\vec{n}}|^2
e^{ - E_{\vec{n}} \tau}
+
\left( 
g_A + \frac{25}{81} g_{\D \D}
\right)
|Z_{\vec{n}}^\D|^2
e^{ - (E_{\vec{n}} + \D) \tau}
\right]
,\end{eqnarray}
from which we can deduce the two couplings 
\begin{equation}
\mathfrak{g}_{\pi N, \pi N} 
= 
- \frac{8}{9} g_A, 
\quad
\text{and}
\quad
\mathfrak{g}_{\pi \D, \pi \D}
=
-
\left( 
g_A + \frac{25}{81} g_{\D \D}
\right)
.\end{equation}
These couplings are momentum independent.

Finally, 
the one-loop infinite volume result contains
$\pi N$-to-$\pi \D$
couplings. 
These appear in the expression
\begin{equation}
G_A^{\pi N, \pi \D} (\tau, t)
=
\frac{16 g_{\D N}}{ 27 \sqrt{2}}
\int_{m_\pi}^\infty
dE 
\sqrt{\ol \rho_{\pi N} (E)
\ol \rho_{\pi \D} (E+ \D)}
\left[
e^{ - E (\tau-t)} e^{- (E + \D) t}
+
e^{ - (E + \D) ( \tau -t)} e^{ - E t}
\right]
.\end{equation}
The corresponding result on a spatial torus is immediate utilizing 
Eq.~\eqref{eq:replacement}
\begin{equation}
G_A^{\pi N, \pi \D} (\tau, t)
=
\frac{16 g_{\D N}}{ 27 \sqrt{2}}
\sum_{\vec{n}}
|Z_{\vec{n}}| \, |Z_{\vec{n}}^\D|
\left[
e^{ - E_{\vec{n}} (\tau-t)} e^{- (E_{\vec{n}} + \D) t}
+
e^{ - (E_{\vec{n}} + \D) ( \tau -t)} e^{ - E_{\vec{n}} t}
\right]
,\end{equation}
and allows us to identify the final coupling, 
which is also momentum independent
\begin{equation}
\mathfrak{g}_{\pi N, \pi \D}
= 
\frac{16}{ 27 \sqrt{2}} g_{\D N}
.\end{equation}

%%%%%%%%%%%%%%%%%%
\subsection{Axial Correlator Results}%
\label{sec:ACR}
%%%%%%%%%%%%%%%%%%

%%%%%%%%%%%%%%%%%%%%%%%%%%%%%%%%%%%%%%%
%
%
%
\begin{figure}
\includegraphics[width=0.5\textwidth]{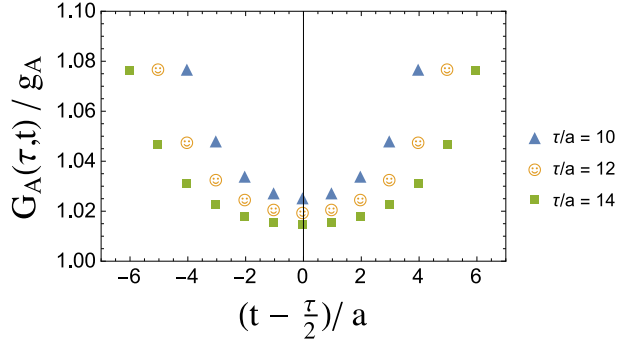}%
\caption{
Excited-state contamination in the axial-vector three-point function of the nucleon,  
Eq.~\eqref{eq:GA}. 
The contamination function 
$G_A ( \tau, t)$
is plotted as a function of the current insertion time $t$ 
(measured relative to the midpoint between source and sink), 
for three values of the source-sink separation, 
$\tau / a = 10$, $12$, and $14$. 
Excited states with pions tend to drive the axial correlator upwards, 
which could potentially lead to an overestimation of 
$g_A$
of a few percent at most.}
\label{f:GA}
\end{figure}
%
%
%
%%%%%%%%%%%%%%%%%%%%%%%%%%%%%%%%%%%%%%%

We have determined the excited-state contributions to the three-point function of the axial-vector current in the nucleon, 
and thereby found the corresponding multi-particle couplings. 
Using the expressions presented above, 
we can write the excited-state contamination in the form
\begin{eqnarray}
G_A (\tau, t)
=
G^{N, \pi N}_A 
+
G^{N, \pi \D}_A 
+
G^{\pi N, \pi N}_A 
+
G_A^{\pi N, \pi \D} 
+
G^{\pi \D, \pi \D}_A 
\label{eq:GA}
,\end{eqnarray}
where the ordering of terms is based on their expected size, 
and we have suppressed the dependence on source-sink separation, 
$\tau$, 
and current insertion time, 
$t$, 
on the right-hand side of the equation. 
In Fig.~\ref{f:GA}, 
we investigate the excited-contamination determined from chiral dynamics. 
Parameter values are taken to be the same as we used above for the two-point function:
lattice parameters are 
$a = 0.1 \, \texttt{fm}$, 
and
$L = 48 \, a$;
the pion mass is the physical value, 
and the nucleon-delta mass splitting is
$\D = 0.29 \, \texttt{GeV}$, 
axial couplings are given values
$g_A = 1.25$, 
$g_{\D N} = 1.5$,  
and
$g_{\D \D} = -2.25$. 
While the final coupling has large uncertainties, 
our results are insensitive to the value due to exponential suppression. 
Results shown in the figure are quite similar to those obtained previously in infinite volume~%
\cite{Tiburzi:2009zp}. 
Contributions from intermediate states with pions tend to drive the axial correlator upwards, 
which would lead to an overestimation of the nucleon axial charge if such contributions are 
not accounted for. 
This finding makes the consistent lattice underestimation of the axial charge more mysterious.

%%%%%%%%%%%%%%%%%%
\section{Summary}%
\label{sec:sum}
%%%%%%%%%%%%%%%%%%

We consider the effect of pion-nucleon and pion-delta states on two- and three-point lattice QCD correlation functions. 
From the perspective of chiral dynamics, 
lattice QCD interpolating operators for the nucleon have definite transformation properties under chiral rotations, 
see~%
\cite{Nagata:2008zzc}. 
As a result, 
these operators can be systematically mapped into effective field theory counterparts~%
\cite{Wein:2011ix}. 
At leading order in the low-energy expansion, 
chiral corrections to the correlators appear with coefficients that are fixed due to the pattern of symmetry breaking~%
\cite{Bar:2015zwa}. 
For lattice nucleon interpolating operators transforming as
$(\frac{1}{2}, 0) \oplus (0,\frac{1}{2})$, 
the interpolating field in the effective theory is not only perturbatively close to the heavy-nucleon field 
$N_v$
of chiral perturbation theory, 
the difference generates only tadpole corrections that do not affect multi-particle couplings. 
This justifies the considerations of an earlier study in infinite volume~%
\cite{Tiburzi:2009zp}, 
provided the quark-level smearing of lattice interpolating operators for the nucleon is small on the scale of the pion Compton wavelength.

For the nucleon two-point function in finite volume, 
the derived couplings to intermediate-state pions above are shown to be in accordance with phase space available on a torus. 
While pion-nucleon and pion-delta spectral weights in finite volume vanish less rapidly at threshold compared to infinite volume, 
pion-nucleon contamination in the two-point function is estimated at the few percent level
(a similar estimate is obtained in~\cite{Bar:2015zwa}), 
while pion-delta contamination cannot be discerned. 
For the nucleon three-point function of the axial-vector current, 
we similarly investigate the effect of pion-nucleon and pion-delta contamination at zero momentum transfer. 
This three-point correlation function enables the lattice QCD determination of the nucleon axial charge, 
$g_A$. 
As in infinite volume,  
chiral corrections to the time-dependence of the axial correlator drive it upwards, 
with the dominant corrections arising from nucleon to pion-nucleon transitions. 
Thus for insufficient time between the source and current insertion, 
as well as between the source and sink, 
chiral corrections could potentially lead to a few-percent overestimation of the axial charge. 
This trend is opposite that encountered in lattice QCD calculations, 
and our study shows that the behavior is not likely due to pion-nucleon contamination. 
It would be interesting to study chiral contamination in other quantities determined from three-point functions, 
such as the quark momentum fraction in the nucleon, 
as results could help lattice practitioners better isolate the single-nucleon contribution.

%%%%%%%%%%%%%%%%%%%%%%%%%%%%%%%%%%%%%%%%%%%%%%%%%%     
\begin{acknowledgments}
This work is supported in part by a joint City College of New York--RIKEN/Brookhaven Research Center fellowship, 
a grant from the Professional Staff Congress of the CUNY,
and by the U.S.~National Science Foundation, under Grant No.~PHY$12$-$05778$.
We are grateful to  
J.-W.~Lee 
for helpful discussions. 
\end{acknowledgments}
%%%%%%%%%%%%%%%%%%%%%%%%%%%%%%%%%%%%%%%%%%%%%%%%%%

%%%%%%%%%%%
\bibliography{bibfile}%
%%%%%%%%%%%

%%%%%%%%%%%
%%%%%%%%%%
%%%%%%%%%
\end{document}